\documentclass[12pt]{article}
\pdfoutput=1
\usepackage{amssymb, amsmath,amsfonts}
\usepackage{multirow}
\usepackage{mathrsfs}
\usepackage{array}
\usepackage{cite}
\usepackage{booktabs}
\usepackage{tikz}
\usepackage{pgfplots}
\usepackage{float}
\usepackage{subfigure, graphicx}
\usepackage[pdftex, bookmarks=true,colorlinks,linkcolor=red,urlcolor=blue,citecolor=blue]{hyperref}
\usepackage{cite}
\usepackage{enumerate}
\usepackage{slashed}
\usepackage{tabularx}
\usepackage{enumerate} 
\usepackage[thicklines]{cancel}

\usepackage{tikz}

\makeatletter\@addtoreset{equation}{section}\makeatother

\def\be{\begin{equation}}
\def\ee{\end{equation}}
\def\bea{\begin{eqnarray}}
\def\eea{\end{eqnarray}}
\newcommand{\comment}[1]{{\bf {\textcolor{blue}{ [#1]}}}}

\def\Dslash{\,\,{\raise.15ex\hbox{/}\mkern-12mu D}}
\def\Dbarslash{\,\,{\raise.15ex\hbox{/}\mkern-12mu {\bar D}}}
\def\delslash{\,\,{\raise.15ex\hbox{/}\mkern-9mu \partial}}
\def\delbarslash{\,\,{\raise.15ex\hbox{/}\mkern-9mu {\bar\partial}}}
\def\pslash{\,\,{\raise.15ex\hbox{/}\mkern-9mu p}}
\def\calDslash{\,\,{\raise.15ex\hbox{/}\mkern-12mu {\cal D}}}

\makeatletter\@addtoreset{equation}{section}\makeatother

\renewcommand{\title}[1]{\vbox{\center\LARGE{#1}}\vspace{5mm}}
\renewcommand{\author}[1]{\vbox{\center#1}\vspace{5mm}}
\newcommand{\address}[1]{\vbox{\center\em#1}}

\def\arXiv#1{\href{http://arxiv.org/abs/#1}{arXiv:#1}}
\def\arXiv#1#2{\href{http://arxiv.org/abs/#1}{arXiv:#1}}

\def\doi#1#2{\href{http://doi.org/#1}{#2}}

\begin{document}

\unitlength = .8mm

\begin{titlepage}
\vspace{.5cm}
 
\begin{center}
\hfill \\
\hfill \\
\vskip 1cm

\title{Holographic Schwinger-Keldysh effective field theories including a non-hydrodynamic mode}
\vskip 0.5cm

{Yan Liu$^{\,a}$}\footnote{Email: {\tt yanliu@buaa.edu.cn}},  
{~Ya-Wen Sun$^{\,b}$}\footnote{Email: {\tt yawen.sun@ucas.ac.cn}},
{~Xin-Meng Wu$^{\,c}$}\footnote{Email: {\tt xinmeng.wu@sjtu.edu.cn}}
\address{${}^{a}$Center for Gravitational Physics, Department of Space Science,\\ 
and Peng Huanwu Collaborative Center for Research and Education,\\
Beihang University, Beijing 100191, China}


\address{${}^{b}$School of Physical Sciences, University of Chinese Academy of Sciences,\\ Beijing 100049, China}

\address{${}^{c}$School of Physics and Astronomy, Shanghai Jiao Tong University, \\Shanghai 200240, China}

\end{center}
\vskip 0.5cm

\abstract{We derive the Schwinger-Keldysh effective field theories for diffusion including the lowest non-hydrodynamic degree of freedom from holographic Gubser-Rocha systems. At low temperature the dynamical non-hydrodynamic mode could be either an IR mode or a slow mode, which is related to IR quantum critical excitations or encodes the information of all energy scales. This additional dynamical vector mode could be viewed as an ultraviolet sector of the diffusive hydrodynamic theory. We construct  two different effective actions for each case and discuss their physical properties. In particular we show that the  Kubo-Martin-Schwinger symmetry is preserved. 
}
\vfill

\begin{center}
{\em In memory of Jan Zaanen.}
\end{center}
\end{titlepage}

\begingroup 
\hypersetup{linkcolor=black}
\tableofcontents
\endgroup





\section{Introduction}
One of the pioneering contributions in AdS/CMT by Jan and collaborators was the discovery of a  Fermi surface in strongly correlated fermionic system at finite density  described through holographic duality \cite{Cubrovic:2009ye}. Together with the works \cite{Liu:2009dm, Faulkner:2009wj}, this led to the development of the so-called Leiden-MIT fermion within the study of holographic fermions. 
Since then, 
the semi-holographic approach in holography \cite{Faulkner:2009wj, Faulkner:2010tq} has attracted lots of research interest. It provides a powerful low energy effective field theory (EFT) for  strongly coupled fermionic systems, even though holographic field theory is a theory for all energy scales \cite{Zaanen:2015oix, Zaanen:2021llz, Hartnoll:2016apf}. Partially motivated by this discovery at zero temperature,  remarkable progress has been made to derive the low energy effective theory for holographic liquids at finite temperature in  \cite{Nickel:2010pr}. Further development to construct the effective field theory for holographic systems can be found in \cite{Crossley:2015tka, deBoer:2015ija} and \cite{Glorioso:2018mmw, deBoer:2018qqm}.  

Hydrodynamics is a universal effective field theory of many-body systems at finite temperature. 
The separation of characteristic energy scales is crucial to build a theory for hydrodynamics of a quantum many-body system. At sufficient late time and long distances, i.e. in the hydrodynamic regime, the only excitations that survive are the  ``slowest" degrees of freedom. However, there are important physical situations that the systems contain extra modes which do not decay fast and their impact to the late time physics should be considered, including weakly momentum relaxation in condensed matter physics, axial charge relaxation in quark gluon plasma etc., which results in observable physical effects, such as charge density waves for momentum weakly broken systems \cite{Baggioli:2022pyb} and negative magneto-resistivity in chiral anomalous  systems \cite{Landsteiner:2014vua}.  

A breakthrough on hydrodynamics is to go beyond the equations of motion to the low energy effective action \cite{Crossley:2015evo, Haehl:2015uoc},  
which is formulated 
from path integrals on a Schwinger-Keldysh (SK) contour from principles of symmetry and unitarity. 
In the SK effective action, the effects of stochastic   fluctuations from random noises and quantum loop corrections can be incorporated naturally. For example, 
the phenomenological entropy current condition in hydrodynamics can also be derived from the SK effective actions \cite{Glorioso:2016gsa}. The nontrivial corrections to the thermal diffusion coefficient due to the thermal fluctuations have been found in \cite{Chen-Lin:2018kfl}.  New stochastic transport coefficients are identified in \cite{Jain:2020zhu} due to the stochastic interactions in the framework of SK effective field theory. 

One remarkable insight is, the hydrodynamic description can be applied to the metallic system that is far from the standard Fermi liquid theory \cite{Davison:2013txa}. This is essentially different from the Fermi liquid state where the quasiparticles interact weakly with high viscosity to avoid the local equilibrium, which makes the hydrodynamics not applicable. 
The idea is that the low energy theory of strange metal can be described by a strongly interacting locally quantum critical state with minimal viscosity coupled to random disorder, which behaves like a hydrodynamic liquid and produces a resistivity proportional the electronic entropy \cite{Davison:2013txa, Balm:2022bju}. The holographic Gubser-Rocha model could have a nice property that the entropy is proportional to temperature at low temperature and could serve as a candidate theory for strange metals \cite{Davison:2013txa}. 
Furthermore, possible turbulence on nanoscale for electron fluid is predicted based on this mechanism  \cite{Zaanen:2018edk}. Further discussions on linear resistivity and other transports in the Gubser-Rocha model can be found in \cite{Jeong:2018tua, Jeong:2021wiu, Balm:2022bju, Ahn:2023ciq}.

From the developments above, it is expected that the hydrodynamics aspects of the Gubser-Rocha model play  important roles in understanding the physics of strange metals and deserves to be further explored. The breakdown of hydrodynamics for the dual system of the Gubser-Rocha model due to the collisions between the hydrodynamic mode and the lowest non-hydrodynamic mode has been studied in \cite{Liu:2021qmt}. At low temperature it was found that  there are two possible different non-hydrodynamic modes, either the IR mode or the slow mode, resulting in different physical behaviors on  transports. 

The aim of this study is to construct the effective field theory for fluctuating hydrodynamics including the dynamical  non-hydrodynamic mode at the tree level from the holographic Gubser-Rocha model. 
We will use the holographic prescription in 
\cite{Glorioso:2018mmw} to complexify the gravitational spacetime and study the probe fields along a specific radial-contour in the bulk. We will introduce a matching procedure on the contour near the horizon. For the slow mode case, The matching point is chosen to be at a distance from the horizon that is sufficiently small in comparison to the temperature. We will see that the resulting effective action matches with the one studied in \cite{Jain:2023obu} for the Maxwell-Cattaneo model of diffusion. For the IR mode case, the matching point is chosen to be at a distance from the horizon that is small compared to the strength of the disorder. We will give a new action from this procedure. 

The plan of this paper is as follows. In Sec. \ref{sec2} we will review the  Schwinger-Keldysh effective field theory of the Maxwell-Cattaneo model studied in \cite{Jain:2023obu}. In Sec. \ref{sec3} we will study the SK effective actions from the holographic Gubser-Rocha model. In Sec. \ref{sec:cd} we conclude and discuss the open questions. 

\section{The Schwinger-Keldysh effective field theory of the Maxwell-Cattaneo model}
\label{sec2}

To include a non-hydrodynamic mode in the effective field theory of diffusive hydrodynamics, we utilize the Maxwell-Cattaneo (MC) model \cite{Jain:2023obu}, which contains an additional vector mode to ``UV regularize" the ordinary diffusion theory to be causal and stable. It belongs to model C in the Hohenberg-Halperin classification \cite{hohenber}. We will review the Maxwell-Cattaneo model  of diffusive hydrodynamics and the Schwinger-Keldysh effective action for this model \cite{Jain:2023obu}. The final SK effective action can be viewed as an effective field theory for diffusive hydrodynamics including a gapped mode.

Consider a quantum many-body system with a conserved $U(1)$ current $J^\mu$ at finite temperature. The standard current conservation equation is\footnote{We work in most plus signature in this work. We ignore the conservation equation for the energy momentum tensor assuming that this equation is decoupled from the conservation of charge here.}   
\be
\label{eq:conservation}
\partial_\mu J^\mu=0\,,
\ee
and the related constitutive equation is 
\be
J^\mu=  n u^\mu-\sigma T \Delta^{\mu\nu} \partial_\nu (\mu/T) +\chi_T\Delta^{\mu\nu}\partial_\nu T \,,
\ee
where $u^\mu$ is the four-velocity of the fluid element, with $n$ the charge density, $\mu$ the chemical potential, and $T$ the temperature. $\Delta ^{\mu\nu}=\eta^{\mu\nu}+u^\mu u^\nu$, and  $\sigma, \chi_T$ are the electric and thermal conductivities. 

In the Maxwell-Cattaneo model, 
an additional vector mode $v_\mu$ is added to the constitutive equation, which satisfies $v^\mu u_\mu=0$ so that $v^\mu$ has only spatial components when we choose $u^\mu$ to be $(1,0,0)$ in the local rest frame. This shares similarity with the Hydro+ theory where an additional scalar field is added \cite{Stephanov:2017ghc}. Now the first law of thermodynamics is modified to be 
\be
d\epsilon= T ds+ \mu dn +\frac{\chi_v}{2} dv^2
\ee
with a new thermodynamic coefficient $\chi_v$ and $v^2=v^\mu v_\mu$. In equilibrium $v_\mu=0$, then the above system reduces to the usual hydrodynamic theory. The new constitutive equation is 
\be
\label{eq:currentKovtun}
J^\mu= n u^\mu + \alpha_v v^\mu
\ee
where  
\be
\label{eq:constiKovtun}
\alpha_v v^\mu=-\sigma \Delta^{\mu\nu} \Big(T \partial_\nu (\mu/T) 
+\frac{\chi_v}{\alpha_v}\partial_t v_\nu\Big)
\,\ee 
and $\Delta^{\mu\nu}=\eta^{\mu\nu}+u^\mu u^\nu$. Compared to the simple diffusion model, the spatial current relaxes with time and is reflected in the time derivative term in Eq.\eqref{eq:constiKovtun}.\footnote{In  Eq.\eqref{eq:constiKovtun}, the order counting can be understood as $v^i\sim \mathcal{O}(\partial)\sim \tau^{-1}$ and there could be other derivative terms, such as $\tau\partial_t \partial_i \mu$. Here we have set the coefficients of such terms to zero to isolate the effect of $v^i$ following \cite{Jain:2023obu}. } The local entropy production  still holds with this new constitutive equation \cite{Jain:2023obu}. 

For simplicity, we take the temperature $T$ and velocity $u^\mu$ in \eqref{eq:constiKovtun} to be homogeneous and take variation of the charge density $n$ out of equilibrium.  
Choosing $u^\mu=(1,0,0)$, we have the density $j^t=n$ and current $j^i=\alpha_v v^i=-\frac{\sigma}{\chi}\partial_i n-\frac{\sigma\chi_v}{\alpha_v}\partial_t v^i$ where $\chi$ is the charge susceptibility. 
Let us introduce  $\tau=\frac{\sigma\chi_v}{\alpha_v^2}$ which is the relaxation time of the vector $v_i$ and $D=\frac{\sigma}{\chi}$ which is the diffusion constant. Combining these expressions and the conservation equation 
Eq.\eqref{eq:conservation}, we have 
\be
\label{eq:coneq}
\partial_t j^t+ \partial_i j^i=0\,,~~~~~
\partial_t j^i+ \frac{D}{\tau}\partial_i j^t=-\frac{1}{\tau}j^i\,,~~~~~
\ee
from which one obtains that the excitations satisfy the telegraph equation
\bea
i\omega(1-i\omega\tau)-Dk^2=0\,.
\eea

It is obvious that the hydrodynamic diffusive mode and the non-hydrodynamic mode interplay with each other to satisfy the semicircle law. In MC mode, one could view $v^i$ as an additional field which is approximately conserved, i.e. replacing $j^i$ in  \eqref{eq:coneq} by $v^i$,  it recovers the quasihydrodynamic picture  \cite{Grozdanov:2018fic} where the diffusive hydrodynamics coupled to additional weakly broken quantities. 
One might perform a Hodge dual for the current,  then the second equation in \eqref{eq:coneq} can be understood as a breaking of higher form symmetry by dualising $(-j^t/\chi_{\rho\rho},j^i/{\chi_{jj}})$ into $K_{\mu\nu}$ as in \cite{Davison:2022vqh}.  

In MC theory of diffusion, the retarded Green's functions of density and the transverse  current are
\be
G^R_{J^tJ^t}(\omega, k)=\frac{-\sigma k^2}{i\omega (1-i\omega \tau)- Dk^2}\,,~~~~~~G^R_{J^i_\perp J^i_\perp}=\frac{i\omega\sigma}{1-i\omega \tau}
\ee
with the energy gap  $\tau^{-1}=\frac{\alpha_v^2}{\sigma\chi_v}$.

\subsection{The Schwinger-Keldysh effective field theory}

The Schwinger-Keldysh effective action for the Maxwell-Cattaneo model is constructed in  \cite{Jain:2023obu}. 
In the Schwinger-Keldysh formalism, the quantum fields 
time evolves along a closed time path. 
Compared to the diffusive hydrodynamics introduced in \cite{Crossley:2015evo}, additional spatial vector fields $v_{1,2\mu}$ are included which satisfy the 
constraints
$ u^\mu v_{1,2\mu}=0,  $
where $u^\mu$ is the four velocity of the fluid elements. When we choose $u^\mu=(1,0,0)$, we have the double fields $v_{1i}$ and $v_{2i}$.

Starting from the microscopic theory of the quantum many-body system, we could integrate out all the fast decay degrees of freedom to obtain the low energy effective theory. In the SK formalism, the closed time path generating functional is 
\bea
\label{eq:skeftft}
\begin{split}
e^{W[A_{1\mu}, A_{2\mu}]}&=\int\mathcal{D}\varphi_1\mathcal{D}\varphi_2 \,  e^{i I_\text{eff}[B_{1\mu},B_{2\mu}]}\\
&=\int\mathcal{D}\varphi_1\mathcal{D}\varphi_2 \mathcal{D}v_{1i}\mathcal{D} v_{2i} \, e^{i S_\text{eff}[B_{1\mu},B_{2\mu},v_{1i},v_{2i}]}
\end{split}
\eea
where 
\be
B_{1\mu}=A_{1\mu}+\partial_\mu\varphi_1\,,~~~~
B_{2\mu}=A_{2\mu}+\partial_\mu\varphi_2\,
\ee
with $A_{1\mu}, A_{2\mu}$ the external sources of the currents, and $\varphi_1, \varphi_2$ as well as $v_{1i}, v_{2i}$ the dynamical variables. 
$I_\text{eff}[B_{1\mu}, B_{2\mu}]$ is a local effective action after integrating out all the UV modes whose precise form has been introduced in \cite{Crossley:2015evo} and see \cite{Liu:2018kfw, Haehl:2024pqu} for reviews.  $S_\text{eff}[B_{1\mu},B_{2\mu},v_{1 i},v_{2i}]$ is the action that we have included the lowest gapped mode which might plays an important role in the underlying physics depending on the energy scale under consideration. Here the dynamical fields $v_{1 i},v_{2i}$ are independent of $\varphi_1,\varphi_2$.  

This work will focus on its properties and its form in both field theory and holography.   
The dynamical fields  $\varphi_1, \varphi_2, v_1, v_2$ in \eqref{eq:skeftft} satisfy the gluing conditions along the closed time path
\bea
\label{eq:bnd}
\begin{split}
&\lim_{t\to\infty}(\varphi_1(t)-\varphi_2(t))=\lim_{t\to-\infty}(\varphi_1(t)-\varphi_2(t-i\beta))=0\,,~~~\\&
\lim_{t\to\infty}(v_{1i}(t)-v_{2i}(t))=\lim_{t\to-\infty}(v_{1i}(t)-v_{2i}(t-i\beta))=0\,.~~~
\end{split}
\eea
It is convenient to introduce the $r$- and $a$-fields, which are defined as 
\be
A_{r\mu}=\frac{1}{2} (A_{1\mu}+A_{2\mu})\,,~~~
A_{a\mu}= A_{1\mu}- A_{2\mu}\,,~~~
\ee
similarly for the fields $\varphi_a, \varphi_r, v_{ai}, v_{ri}$. The currents are defined as 
\be
J^\mu_r(x)=\frac{\delta I_\text{eff}}{\delta A_{a\mu}(x)}\,,~~~
J^\mu_a(x)=\frac{\delta I_\text{eff}}{\delta A_{r\mu}(x)}\,.
\ee

On each leg of the SK contour, there is a $U(1)$ symmetry, 
\be
\varphi_a\to \varphi_a+\lambda_a\,,~~\varphi_r\to \varphi_r+\lambda_r\,,~~
A_{a\mu}\to A_{a\mu}-\partial_\mu\lambda_a\,,~~
A_{r\mu}\to A_{r\mu}-\partial_\mu\lambda_r\,,
\ee
the variables $B_\mu=A_\mu+\partial_\mu\varphi\,$ are consistent with this symmetry. 
Here the fields $\lambda_a$ and $\lambda_r$ are time dependent, and the above gauge transformation needs to 
preserve the boundary condition \eqref{eq:bnd}. The $U(1)\times U(1)$ symmetry is broken to $U(1)$ for time independent gauge transformations where only the diagonal subgroup preserves the boundary condition \eqref{eq:bnd}.

There are some constraints on $S_\text{eff}$ from symmetry and unitarity \cite{Jain:2023obu}. The following three constraints are from continuous symmetry transformations. 
\begin{itemize}
\item[(1)] translational invariance of time shift;
\item[(2)] rotational invariance among spatial directions;
\item[(3)] Chemical shift symmetry between the two dynamical fields 
\be
\label{eq:chess}
\varphi_r\to \varphi_r+\sigma(x_i)\,,~~~\varphi_a\to \varphi_a
\ee
where $\sigma(x_i)$ is time-independent. 
Equivalently, one can write it as 
\be
B_{a\mu}\to B_{a\mu}\,,~~~B_{r0}\to B_{r0}\,,~~~B_{ri}\to B_{ri}+\partial_i\sigma(x_j) \,,~~~
\ee
which is a time-independent, diagonal gauge transformation. We do not have such symmetry for the dynamical fields $v_{ai}, v_{ri}$. This symmetry indicates that the allowed form $\partial_t B_{ri}$ in the action. 
\end{itemize}

Furthermore, $S_\text{eff}$ should also satisfy the following three unitary conditions \cite{Liu:2018kfw}:
\begin{itemize}
\item[(4)] A $Z_2$ reflection symmetry, i.e. interchange the two contour, 
\be 
(S_\text{eff}[B_{a\mu},B_{r\mu},v_{ai},v_{ri}])^*=-
 S_\text{eff}[-B_{a\mu},B_{r\mu},-v_{ai},v_{ri}]
 \,;
 \ee
 \item[(5)] When we set all the sources and dynamical fields of the two legs to be equal, we have 
\be  S_\text{eff}[B_{a\mu}=0,B_{r\mu},v_{ai}=0,v_{ri}]= 0
\,;
\ee
 \item[(6)] The imaginary part is non-negative; 
\be \text{Im} S_\text{eff}[B_{a\mu},B_{r\mu},v_{ai},v_{ri}]\geq 0\,.
\ee
\end{itemize}

Moreover, there is another discrete symmetry, i.e. 
\begin{itemize}
\item[(7)] When the underlying microscopic theory is CPT invariant, the low energy effective theory is invariant under dynamical Kubo-Martin-Schwinger (KMS) symmetry
\be
S_\text{eff}[\hat{B}_{a\mu},\hat{B}_{r\mu},\hat{v}_{ai},\hat{v}_{ri}]
=S_\text{eff}[B_{a\mu},B_{r\mu},v_{ai},v_{ri}]
\ee
where 
\be
\hat{B}_{r\mu}= \Theta B_{r\mu}\,,~~\hat{v}_{ri}= \Theta v_{ri}\,,~~\hat{B}_{a\mu}= \Theta (B_{a\mu}+i\beta\partial_0 \hat{B}_{a\mu})\,,~~\hat{v}_{ai}= \Theta (v_{ai}+i\beta\partial_0 \hat{v}_{ai})
\ee
with $\Theta$ a discrete C or P or T symmetry or their combinations \cite{Jain:2023obu}. 
\end{itemize}

The 
effective action can be derived from 
either the Martin-Siggia-Rose (MSR) formalism or the
Schwinger-Keldysh formalism.   It takes the following form at the quadratic order of  $B_{a,r}$, and linear order in derivatives of $B_r$ without any explicit spatial derivatives \cite{Jain:2023obu},  
\bea
\label{eq:action-Kovtun}
S_\text{eff}[B_{a\mu},B_{r\mu},v_{ai},v_{ri}]&=&\int d^{d+1} x \bigg[\chi B_{a0}B_{r0}-\chi_v v_{ai}v_{ri}+i T \sigma \Big(B_{ai}+\frac{\chi_v}{\alpha_v}v_{ai}\Big)
\Big(B_{ai}+\frac{\chi_v}{\alpha_v}v_{ai}\Big)\nonumber\\
&&~~-\sigma \Big(B_{ai}+\frac{\chi_v}{\alpha_v}v_{ai}\Big)\partial_t\Big(B_{ri}+\frac{\chi_v}{\alpha_v}v_{ri}\Big)\,.
\eea 
When $\chi_v\to 0$, it reduces to the standard EFT for diffusive hydrodynamics \cite{Crossley:2015evo}
\be
I_\text{eff}[B_{a\mu},B_{r\mu}]=\int d^{d+1} x  \bigg[\chi B_{a0}B_{r0}+i T \sigma B_{ai}^2-\sigma B_{ai}\partial_0 B_{ri}\big)\Big)\bigg]\,.
\ee

\section{Effective field theories from holography}
\label{sec3}
In this section, we study the Schwinger-Keldysh effective field theories from the holographic Gubser-Rocha model.  The holographic Gubser-Rocha model contains two different types of non-hydrodynamic modes at low temperature \cite{Liu:2021qmt}. We will follow the holographic prescription introduced in \cite{Glorioso:2018mmw} to study the Schwinger-Keldysh action. The new ingredient is that when we consider a  cutoff  near IR and perform the matching procedure for the fields at the cutoff, we can obtain the effective action including both the hydrodynamic  mode and the lowest non-hydrodynamic mode. 

In our setup the $U(1)$ symmetry is preserved and an additional dynamical vector field 
associated with the non-hydrodynamic mode is introduced to make the diffusion theory causal and stable. Our study differs from \cite{Baggioli:2023tlc, Hongo:2024brb} where the effective theory for an explicitly weakly broken $U(1)$ conservation was studied. The SK effective action for spontaneously broken $U(1)$ symmetry was studied in \cite{Bu:2021clf, Donos:2023ibv, Bu:2024oyz}. For other developments on SK effective field theory from holography, see e.g. \cite{Jana:2020vyx, Abbasi:2022aao, Abbasi:2024pwz, Knysh:2024asf, Jain:2020hcu, Bu:2020jfo, Ghosh:2020lel, Ghosh:2022fyo,  Mullins:2023ott}.

\subsection{The holographic Gubser-Rocha model}

We first shortly review the holographic Gubser-Rocha mode \cite{Liu:2021qmt}, whose action is 
\be\label{eq:actionbg}
S=\int d^4x\sqrt{-g}\,\bigg(R-\frac{3}{2}(\partial\phi)^2+\frac{6}{L^2}\cosh\phi-\frac{1}{2}\sum_{I=1}^2(\partial\psi_I)^2\bigg)\,.
\ee
The massless scalar fields $\psi_I=m\,x^i\,\delta_{Ii}$ are known as linear axion fields \cite{Andrade:2013gsa} which explicitly break the spatial translational symmetry in the $x$-$y$ plane while preserving the isotropy. It provides an additional energy scale $m$ compared to the temperature $T$. 
The dilaton field $\phi$ with a particular choice of potential is used to realize a special quantum critical point at low energy. 

We focus on the zero density solution with the following form \cite{Liu:2021qmt}
\be
\label{eq:metric}
ds^2=-u dt^2+\frac{dr^2}{u}+f (dx^2+dy^2)\,,~~~\phi=\phi(r)\,,~~~\psi_I=mx_I
\ee
where
\bea
\label{eq:neubg}
\begin{split}
u&=\sqrt{r}(r-r_0)\left(r-r_0+\sqrt{2}m\right)\frac{1}{\sqrt{r-r_0+\frac{m}{\sqrt{2}}}}\,,\\
f&= \sqrt{r}\left(r-r_0+\frac{m}{\sqrt{2}}\right)^{3/2}\,,\\
\phi&= \frac{1}{2}\log\left(\frac{r-r_0+\frac{m}{\sqrt{2}}}{r}\right)\,
\end{split}
\eea
are functions of the radial coordinates $r$ in the bulk with $r_0$ the location of the horizon. Approaching to the UV boundary $r\rightarrow \infty$, the bulk geometry is asymptotic to AdS$_4$ with the negative cosmological constant reflected in the potential of the dilaton. At extremely low temperature, the black hole has a near horizon geometry which is conformal to AdS$_2\,\times$ R$^2$. 

In the ingoing Eddington–Finkelstein (iEF) coordinates  $v=t+\tilde{r}$ with $d\tilde{r}/dr=u^{-1}$, the metric \eqref{eq:metric} becomes  
\be
ds^2=-u dv^2+2 dv dr + f(dx^2+dy^2)\,,~~~
\ee
where $u, f$ and $\phi$ take the same form as \eqref{eq:neubg}. In the following we will use the iEF coordinates to calculate the effective action.

\subsection{Effective field theories}

The action for the bulk gauge field $C_m$ is
\be
S=-\frac{1}{4}\int d^4x \sqrt{-g} \,e^{\alpha\phi}F_{mn}F^{mn}\,,
\ee
where $F_{mn}=\partial_m C_n-\partial_n C_m$ is the field strength. The gauge field couples to the dilaton $\phi$ via a tunable parameter $\alpha$.  This gauge field $C_m$ in the bulk is dual to a conserved current $J^\mu$ at the boundary. From the holographic dictionary, the dual quantities electric conductivity $\sigma$ and susceptibility $\chi$ are 
\bea
\sigma=\left(\frac{m}{2\sqrt{2}\pi T}\right)^\alpha\,,
\quad
\chi=\frac{\left(1+\alpha\right) m^\alpha \left(m^2-8\pi^2 T^2\right)}
 {2\sqrt{2}m^{1+\alpha}-2^{3+\frac{3\alpha}{2}}(\pi T)^{1+\alpha}}\,,
\eea
where we assume $\alpha\geq 0$, and the charge  diffusive constant $D=\sigma/\chi$. Note that at low temperature we have $\chi\sim m$. One can find that the charge diffusion depends on the coupling $e^{\alpha\phi}$, and indeed the parameter controls not only the diffusive hydrodynamics but also the relaxation properties of the lowest non-hydrodynamic excitations. In \cite{Liu:2021qmt}, it was observed that there exits two different types of non-hydrodynamic modes at low temperature, i.e. the IR mode with lifetime $\tau\sim 1/T$ when $\alpha<1$ and the slow mode with lifetime $\tau\sim 1/T^\alpha$ when $\alpha>1$. The pole collision can determine the convergence of dispersion  relation \cite{Grozdanov:2019kge}.  Especially when $\alpha>1$, the behavior of the hydrodynamic mode and the first non-hydrodynamic mode can be characterized by 
the telegrapher equation derived in the appendix of \cite{Liu:2021qmt}
\bea\label{eq:teleq}
\omega^2+\frac{i}{\tau}\omega-\frac{D_c}{\tau}k^2=0\,,
\eea
where 
\bea
\label{eq:Dtau}
\tau=\int_{r_0}^\infty ds\left(\frac{e^{\alpha\phi_0}}{u e^{\alpha\phi}}-e^{\phi_0}\,
\bigg[\frac{1}{s^2-r_0^2}-\frac{\log(s-r_0)}{(s+r_0)^2}\bigg]\right)\,,~~~~
D_c=\int_{r_0}^{\infty}dr\,\frac{e^{\alpha\phi_0}}{f e^{\alpha\phi}}\,
\eea
with $\phi_0=\frac{1}{2}\log\frac{m}{\sqrt{2} r_0}$.


The equation of motion for the gauge field is  
\be
\label{eq:eomCmu}
\text{E}^n=\frac{1}{\sqrt{-g}}\partial_m (e^{\alpha \phi}\sqrt{-g} F^{mn})=0\,.
\ee
For the latter purpose to solve the EOM order by order in $\partial_\mu$ perturbatively, let's define 
\be \Pi^\mu=\sqrt{-g} e^{\alpha\phi}F^{\mu r}\,,
\ee
and we have
\bea
\label{defPi}
\begin{split}
\Pi^0&=f e^{\alpha\phi}\,\partial_r C_0\,,\\
\Pi^i&=- e^{\alpha\phi}(u\partial_r C_i+\partial_0 C_i-\partial_i C_0)\,.
\end{split}
\eea
Therefore, the EOM $\text{E}^\mu=0$ becomes
\bea
\label{eomPi}
\begin{split}
\partial_r\Pi^0&=-\partial_i(e^{\alpha\phi}\partial_r C_i)\,,\\
\partial_r\Pi^i&= \partial_0(e^{\alpha\phi}\partial_r C_i)+f^{-1} e^{\alpha\phi}  \partial_j F_{ji} \,.
\end{split}
\eea
Due to the Bianchi identity $\partial_r(\sqrt{-g}E^r)+\partial_0(\sqrt{-g}E^0)+\partial_i(\sqrt{-g}E^i)=0$, we only need to impose $E^r=0$ at a single slice of $r$, e.g. we could choose $\lim_{r\to\infty}E^r=0$ which is the conservation of currents.  
The on-shell action can be written as a boundary term
\bea
\label{eq:onshellS}
\begin{split}
S&=-\frac{1}{4} \int d^4x \sqrt{-g}\, e^{\alpha\phi} F_{mn} F^{mn}\\
&=-\frac{1}{2}\int d^4x\sqrt{-g}\,\bigg[\frac{1}{\sqrt{-g}}\partial_m\left(e^{\alpha\phi}\sqrt{-g} F^{mn}C_n\right)-\text{E}^n\cdot C_n\bigg]\\
&=\frac{1}{2}\int dvdxdy\,
\left(-e^{\alpha\phi}\sqrt{-g} F^{rn}C_n\right)\Big{|}_{\text{IR}}^{\text{UV}},
\end{split}
\eea
where we have used the EOM \eqref{eq:eomCmu} and we assume the total derivative only contributes at the $r$-slices. For example, if we assume all the contributions are from the UV boundary and focus on the hydrodynamic limit, we obtain the EFT for charge diffusion \cite{Glorioso:2018mmw}.

\begin{figure}[h]
\begin{center}
\begin{tikzpicture}
\node at (9.5, 1.45) {$\infty_2$};
\node at (9.5, -0.1) {$\infty_1$};
\node[color=red] at (6.3,-0.4) {$\mathcal{C}_1$};
\node[color=red] at (-1.3,0.7) {$\mathcal{C}_2$};
\node[color=red] at (6.3,1.8) {$\mathcal{C}_3$};
\node[color=blue] at (2.7,-0.3) {$r_*-i\epsilon$};
\node[color=blue] at (2.7,1.7) {$r_*+i\epsilon$};
\node[color=gray] at (0.1, 0.4) {$r_0$};
\draw[thick] (1,1.4) arc (40:320:1.1cm);
\draw[thick][color=black] [->] (-0.93,0.73) -- (-0.93,0.65);
\draw[thick][color=black] [->] (1,0) -- (6,0);
\draw[thick][color=black](6,0) -- (9,0);
\draw[color=black] [->] (0.1,0.7) -- (9.3,0.7);
\draw[thick][color=black] [->]  (9,1.4)--(6.8,1.4);
\draw[thick][color=black] (6.8,1.4) -- (1,1.4);
\draw[dashed, color=blue] (2,2.4)--(2,-1);
\end{tikzpicture}

\end{center}
\vspace{-0.3cm}
\caption{\small The radial contour in the bulk which is dual to Schwinger-Keldysh contour in the field theory. Here we complexify the bulk geometry and consider the gravitational system in the complex $r$-plane.  We will choose $\omega, k\ll r_*-r_0\ll T$ for the slow mode while $\omega, k\ll r_*-r_0\ll m$ for the IR mode.
}
\label{fig:contour}
\end{figure}
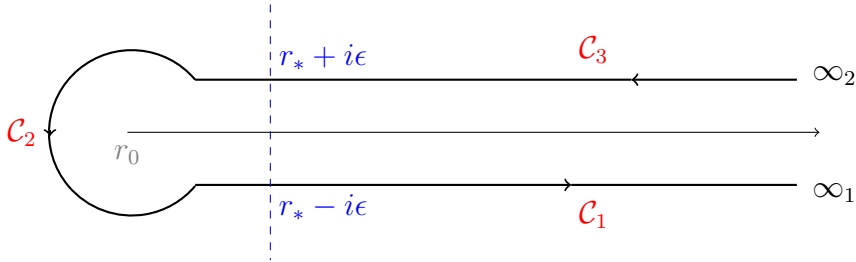

We choose the bulk contour which is dual to the Schwinger-Keldysh contour as follows. First, we follow the strategy in \cite{Glorioso:2018mmw} to complexify the gravitational spacetime via generalizing the radial coordinate $r$ into the complex plane with a small imaginary part. Different from \cite{Glorioso:2018mmw}, we introduce an IR cutoff at $r_*$ to divide the contour $\mathcal{C}$ into three parts, i.e. $\mathcal{C}=\mathcal{C}_1\cup\mathcal{C}_2\cup\mathcal{C}_3$, where $\mathcal{C}_1$ and $\mathcal{C}_3$ are the far regions, while $\mathcal{C}_2$ is the near region.\footnote{Similar cutoff has been used in \cite{deBoer:2018qqm, Daguerre:2023cyx}.} The far region is always chosen to be $r-r_0\gg\omega, k$, in order that we can solve the gauge field $C_m$ order by order in $\omega$ and $k$. The definition for the near region depends on the low-energy physics in consideration. When a slow relaxation mode exist in the system, the near region is $r-r_0\ll T$ with the overlap region defined as $\omega,k \ll r-r_0\ll T$. When the IR excitations are the lowest gapped modes in the system, we choose $r-r_0\ll m$ for the near region to include the whole IR geometries. The overlap region is properly defined as $\omega, k\ll r-r_0\ll m$. Nevertheless, the relations among $T, m$ and $\omega, k$ are still undetermined. We illustrate the separation in the cartoon shown in Fig. \ref{fig:contour}. 

Before moving forward to compute the effective action, we make explanations for the boundary conditions and their roles in the dual field theories. 
We impose the Dirichlet boundary conditions at infinity as 
\be
C_\mu (r\to\infty_1)=A_{1\mu}\,,~~~~~~C_\mu (r\to\infty_2)=A_{2\mu}\,,~~~
\ee
where $\infty_{1}=\infty-i\epsilon,  \infty_{2}=\infty+i\epsilon,$ and $A_{1\mu}, A_{2\mu}$ are sources of $J_1^\mu, J_2^\mu$. 
We also impose Dirichlet boundary conditions at the IR cutoff, i.e.,
\be
C_\mu (r\to r_{* 1})=a_{1\mu}\,,~~~~~~C_\mu (r\to r_{*  2})=a_{2\mu}
\ee
where $r_{*  1}=r_* -i\epsilon, r_{*  2}=r_* +i\epsilon$. Notice that $A_{1\mu},A_{2\mu},a_{1\mu},a_{2\mu}$ are constants in equilibrium and can still depend on $x^\mu$ for near-equilibrium. 


From gauge transformations $C_m\to C_m+\partial_m \Lambda$,  we can set the gauge $C_r=0$ by choosing 
\be
\Lambda(r,x^\mu)=\int_r^{r_0} dr' \, C_r(r',x^\mu)\,.
\ee 
Here $r_0$ is a reference point and could be understood as a point on the circle 
in Fig. \ref{fig:contour} in the limit $\epsilon\to 0$ \cite{Glorioso:2018mmw}. 
The boundary conditions are
\be
C_\mu(r\to\infty_1)=B_{1\mu}=A_{1\mu}+\partial_\mu \varphi_1\,,~~~~
C_\mu(r\to\infty_2)=B_{2\mu}=A_{2\mu}+\partial_\mu \varphi_2\,,
\ee
where 
\be
\varphi_1=\int_{\infty_1}^{r_0}dr\, C_r\,,~~~~
\varphi_2=\int_{\infty_2}^{r_0}dr\, C_r\,,
\ee
are the Wilson line degrees of freedom associated to $C_r$. 
In the gauge $C_r= 0$, there is a residual gauge
\be
C_\mu \to C_\mu +\partial_\mu \lambda(x^\mu)
\ee
which changes $\varphi_1\to \varphi_1+\lambda, \varphi_2\to \varphi_2+\lambda$. Thus $\varphi_1$ 
and $\varphi_2$ are not independent. At the cutoff $r_*$, the boundary condition becomes 
\be
C_\mu (r\to r_{* 1})=b_{1\mu}=a_{1\mu}+\partial_\mu\int_{r_{*1}}^{r_0}dr\, C_r\,,~~~C_\mu (r\to r_{*  2})=b_{2\mu}=a_{2\mu}+\partial_\mu\int_{r_{*2}}^{r_0}dr\,C_r\,.
\ee

Furthermore, we follow \cite{Glorioso:2018mmw} to impose
the boundary condition 
\be \label{eq:c0bnd}
C_0(r_0)=0\,.
\ee 
With the condition above, there exists a residual gauge transformation which is precisely the shift symmetry introduced in \eqref{eq:chess}. 

Here we assume that the infrared fields $b_{1\mu}$ and $b_{2\mu}$ are dynamical and this is different from the setup in \cite{Glorioso:2018mmw} where these two gauge fields have been integrated out at the beginning in \cite{Glorioso:2018mmw}. 
Similar to the physical picture introduced in \eqref{eq:skeftft}, the effective action has the following form, 
\bea
\begin{split}
e^{W[A_{1\mu}, A_{2\mu}]}&=\int\mathcal{D}\phi_1\mathcal{D}\phi_2   e^{i I_\text{eff}[B_1,B_2]}\nonumber\\
&=\int\mathcal{D}\phi_1\mathcal{D}\phi_2 \mathcal{D}b_{1\mu}\mathcal{D} b_{2\mu}  e^{i S_\text{eff}[B_1,B_2,b_{1\mu},b_{2\mu}]},
\end{split}
\eea
where the effective action is 
\be
S_\text{eff}[B_{1\mu}, B_{2\mu},b_{1\mu},b_{2\mu}]=
S_{\mathcal{C}_1\cup\mathcal{C}_3}[b_{1\mu}, B_{1\mu}, b_{2\mu}, B_{2\mu}]+S_{\mathcal{C}_2}[b_{1\mu}, b_{2\mu}]
\,,
\ee
with $
B_\mu=A_\mu+\partial_\mu\varphi$ and 
$\varphi$ is the hydrodynamic field while $b_\mu$ is an emergent dynamical gauge field. The precise relation between $b_\mu$ here and $v_i$ in \eqref{eq:skeftft} will be discussed later. 

\subsubsection{EFT along $\mathcal{C}_1$ and $\mathcal{C}_3$}

The far regions $\mathcal{C}_1$ and $\mathcal{C}_3$ are defined as $\omega, k \ll r-r_0$ along which one can 
perform a derivative expansion for the fields as
\be
C_\mu=C_\mu^{(0)}+C_\mu^{(1)}+C_\mu^{(2)}+\dots
\ee
where the upper index indicates expansion to the $i$-th order in the boundary spacetime derivatives. The boundary conditions should also be imposed order by order, and we always take the source to be the leading order, i.e., 
\bea
\label{eq:bndcmu}
\begin{split}
C_\mu^{(0)} (r\to\infty_1)=B_{1\mu}(x)\,, ~~~~~~~~~~~~~~~~~~~~~~~~C_\mu^{(0)} (r\to\infty_2)=B_{2\mu}(x)\,,&&~~~\\
C_\mu^{(0)} (r\to r_*-i\epsilon)=b_{1 \mu}(x)\,,~~~~~~~~~~~~~~~~~~~C_\mu^{(0)} (r\to r_*+i\epsilon)=b_{2\mu }(x)\,,&&~~~
\end{split}
\eea
while it vanishes in the higher order
\bea
\begin{split}
C_\mu^{(n)} (r\to\infty_1)&=C_\mu^{(n)} (r\to r_*-i\epsilon)=0\,,\\[6pt]
C_\mu^{(n)} (r\to\infty_2)&=C_\mu^{(n)}(r\to r_*+i\epsilon)=0\,
\end{split}
\eea
for $n\geq1$.

To the zeroth order in $\partial_\mu^{(0)}$, the radial partial derivative $\partial_r$ dominates, leading to the equations at zeroth order
\be
\partial_r(f e^{\alpha\phi} \partial_r C_0^{(0)})=0\,,~~~~~
\partial_r(u e^{\alpha\phi} \partial_r C_i^{(0)})=0\,.~~~
\ee
We have the solution of $C_0^{(0)}$ satisfying the boundary condition \eqref{eq:bndcmu} 
\be
C_0^{(0)}=
 \begin{cases}
   -\displaystyle \frac{B_{10}-b_{10}}{Q_0} \displaystyle \int_r^{\infty_1}\frac{d\tilde{r}}{f\, e^{\alpha\phi} }+ B_{10}(x)   \,~~~~~~\text{on~} \mathcal{C}_1  \\[10pt]
  -\displaystyle \frac{B_{20}-b_{20}}{Q_0} \displaystyle \int_r^{\infty_2}\frac{d\tilde{r}}{f\, e^{\alpha\phi} } +B_{20}(x)   \,~~~~~~\text{on~} \mathcal{C}_3
  \end{cases}
\ee
where 
\be
Q_0=\int_{r_*}^\infty\frac{dr}{f\, e^{\alpha\phi} }
\ee
which is regular when $r_{*}\to r_0$.

On the other hand, the solution of $C_i^{(0)}$ takes the form 
\be
C_i^{(0)}=
 \begin{cases}
  - \displaystyle\frac{B_{1i}-b_{1i}}{Q_{i}} \displaystyle\int_r^{\infty_1}\frac{d\tilde{r}}{u\, e^{\alpha\phi}}+ B_{1i}(x)   \,~~~\text{on~} \mathcal{C}_1    \\[10pt]
   - \displaystyle\frac{B_{2i}-b_{2i}}{Q_{i}}\displaystyle\int_r^{\infty_2}\frac{d\tilde{r}}{u\, e^{\alpha\phi}} +B_{2i}(x)  \,~~~\text{on~} \mathcal{C}_3
  \end{cases}
\ee
where 
\be
Q_{i}=\int_{r_*}^{\infty} \frac{dr}{u\, e^{\alpha\phi}}\,.
\ee
There is a logarithmic divergence as $r_*\rightarrow r_0$. However, we restrict the solution to the far region $\omega, k \ll r_*-r_0$, and it is always finite. 
In the following, we will further take $r_*-r_0\ll T$ for $\alpha>1$, and find that $Q_i$ 
it is nothing but $\tau/\sigma$ in this scenario. For $\alpha<1$ we will take $r_*-r_0\ll m$.  

The equations for the fields at higher orders are 
\be
\label{Eqhigher}
\partial_r(f e^{\alpha\phi}\, \partial_r C_0^{(n)})=s_0^{(n)}\,,~~~~~~~~~~~~
\partial_r(u e^{\alpha\phi}\, \partial_r C_i^{(n)})=s_i^{(n)}\,,~~~~~~~~~~~~
\ee
where
\bea
\label{defsource}
\begin{split}
s_0^{(n)}&=-e^{\alpha\phi}\partial_i\partial_r C_i^{(n-1)}\,,\\
s_i^{(n)}&=- e^{\alpha\phi} \partial_0\partial_r C_i^{(n-1)}-
\partial_r\big(e^{\alpha\phi}(\partial_0 C_i^{(n-1)}-\partial_i C_0^{(n-1)})\big)
\\
&~~~~-f^{-1} e^{\alpha\phi}  \partial_j \big(\partial_jC_{i}^{(n-2)}-\partial_i C_j^{(n-2)}\big) \,.
\end{split}
\eea
The source of $s^{(n)}$ is determined 
by lower order solutions. 
For example, when $n=1$ we have  
\bea
s_0^{(1)}&=&
 \begin{cases}
   \displaystyle\frac{\partial_i (B_{1i}- b_{1i})}{Q_{i}}\,\frac{1}{u}\,,  ~~~~~~~~~~~~~~~~~~~~~~~~~~~~~~~~~~~~~~\text{on~} \mathcal{C}_1    \\[8pt]
   \displaystyle\frac{\partial_i (B_{2i}-\partial_i b_{2i})}{Q_{i}} \frac{1}{u} \,,~~~~~~~~~~~~~~~~~~~~~~~~~~~~~~~~~~~~\text{on~} \mathcal{C}_3
    \end{cases}\,,\\[8pt]
~~s_i^{(1)}&=&
\begin{cases}
  \displaystyle \frac{\partial_0( B_{1i}- b_{1i})}{Q_{i}\, }\frac{1}{u}-\partial_r\big(e^{\alpha\phi}(\partial_0 C_i^{(0)}-\partial_i C_0^{(0)})\big)\,,  ~~~\text{on~} \mathcal{C}_1    \\[8pt]
  \displaystyle \frac{\partial_0 (B_{2i}- b_{2i})}{Q_{i}\, }\frac{1}{u}-\partial_r\big(e^{\alpha\phi}(\partial_0 C_i^{(0)}-\partial_i C_0^{(0)})\big) \,,~~~\text{on~} \mathcal{C}_3
    \end{cases}\,.
\eea

From the above equations, the solutions at the first order are
\be
C_0^{(1)}=\int_{\infty}^r dr'\frac{1}{fe^{\alpha\phi}}\bigg[\int_{\infty}^{r'} dr''s_0^{(1)}+c_0^{(1)}\bigg]\,,
\ee
and 
\be
C_i^{(1)}=\int_{\infty}^r dr'\frac{1}{u e^{\alpha\phi}}\bigg[\int_{\infty}^{r'} dr''s_i^{(1)}+c_i^{(1)}\bigg]\,,
\ee
with the integration constants
\bea
\begin{split}
c_0^{(1)}&=\frac{1}{Q_0}\int_\infty^{r_*}dr'\frac{1}{fe^{\alpha\phi}}\int^{r'}_\infty dr'' s_0^{(1)}=
\frac{\partial_i( B_{i}- b_{i})}{Q_0Q_i}\int_\infty^{r_*}dr'\frac{1}{fe^{\alpha\phi}}\int^{r'}_\infty dr'' \frac{1}{u}
\,,\\
c_i^{(1)}&=\frac{1}{Q_i}\int_\infty^{r_*}dr'\frac{1}{ue^{\alpha\phi}}\int^{r'}_\infty dr''s_i^{(1)}\\
&=\frac{\partial_0 (B_{i}- b_{i})}{Q_i^2}\int_\infty^{r_*}dr'\frac{1}{ue^{\alpha\phi}}\int^{r'}_\infty dr''\frac{1}{u}
-\frac{1}{Q_i}\int_\infty^{r_*}dr'\frac{1}{u}\left[\left(\partial_0 C_i^{(0)}-\partial_i C_0^{(0)}\right)-
\left(\partial_0 B_i-\partial_i B_0\right)
\right]\,.
\end{split}
\eea
Note that $c_0^{(1)}, c_i^{(1)}$ can only be computed numerically. We will 
keep these terms in the current form in the on-shell action. 

With the solutions to the first order in the far region, we then calculate the on-shell action along $\mathcal{C}_1$ and $\mathcal{C}_3$. 
The on-shell action can be written as a boundary term as shown in Eq.\eqref{eq:onshellS}, therefore 
\bea
\begin{split}
S_{\mathcal{C}_1\cup\mathcal{C}_3}=\frac{1}{2}\int dvdxdy\,\left[
\left(-e^{\alpha\phi}\sqrt{-g} F^{rn}C_n\right)\Big{|}_{\infty_2}^{r_*+i\epsilon}+
\left(-e^{\alpha\phi}\sqrt{-g} F^{rn}C_n\right)\Big{|}^{\infty_1}_{r_*-i\epsilon}
\right]\,.
\end{split}
\eea
Explicitly, the Lagrangian density in the field theory is 
\bea
\begin{split}
-e^{\alpha\phi}\sqrt{-g}\, F^{rn}C_n
&=e^{\alpha\phi}f\,C_0\partial_rC_0-e^{\alpha\phi}u\,C_i\partial_rC_i+e^{\alpha\phi}C_i\partial_iC_0
\,.
\end{split}
\eea
Note we have used the fact that the integration for  $e^{\alpha\phi}C_i\partial_0C_i$ vanishes. 

Consider $e^{\alpha\phi}f\,C_0\partial_rC_0$ at the order of $\partial_\mu^{(1)}$, 
\bea
e^{\alpha\phi}f\,\left(C_0^{(0)}\partial_rC_0^{(1)}+C_0^{(1)}\partial_rC_0^{(0)}\right)=
\begin{cases}
 B_0 c_0^{(1)}\,, \quad~~~~~~~~~~~~~~~~~~~~~~~ \text{at UV}  \\[8pt]
 b_0\left(\displaystyle\int_\infty^{r_*} dr\,s_0^{(1)}+c_0^{(1)}\right)\,,\quad~~~ \text{at IR}
\end{cases}
\eea
and we find that there is no $\partial_0$ involved since $c_0^{(1)}$ and $s_0^{(1)}$ only involves $\partial_i$. 
Then we consider $-e^{\alpha\phi}u\,C_i\partial_rC_i$ of  $\partial_0^{(1)}$
\bea
-e^{\alpha\phi}u\,\left(C_i^{(0)}\partial_rC_i^{(1)}+C_i^{(1)}\partial_rC_i^{(0)}\right)=
\begin{cases}
 -B_i c_i^{(1)}\,, \quad~~~~~~~~~~~~~~~~~~~~· \text{at UV}  \\[8pt]
 -b_i\left(\displaystyle\int_\infty^{r_*}dr\, s_i^{(1)}+c_i^{(1)}\right)\,,\quad \text{at IR}
\end{cases}
\eea
and this term contributes to the effective action involving time derivatives. 
Finally, $e^{\alpha\phi}\,(C_i^{(0)}\partial_iC_0^{(1)}+C_i^{(1)}\partial_iC_0^{(0)})$ always vanishes at the UV and IR boundaries as we have imposed the vanishing Dirichlet boundary conditions in Eq. \eqref{eq:bndcmu} for the first order solutions of $C_\mu^{(1)}$.  

Plugging in the zeroth order solution and focusing on the diffusive dynamics, we only consider the terms up to $\partial_0^{(1)}$ and ignore the terms involving $\partial_i$, to obtain
\bea
\label{EFTC1C3}
\begin{split}
S_{\mathcal{C}_1\cup\mathcal{C}_3}
&=\int dvdxdy\,\Big[
c_0 (B_{a0}-b_{a0})(B_{r0}-b_{r0})
- c_1 ({B}_{ai}-b_{ai})({B}_{ri}-b_{ri})
\\&~~~~~~~~~~~~~+ 
c_2 (b_{ai}\partial_0 B_{ri}-B_{ai}\partial_0 b_{ri})+\dots
\Big]\,,
\end{split}
\eea
where
\be
c_0=\frac{1}{Q_0}=\chi\,, ~~~~ c_1=\frac{1}{Q_i}\,,~~~~~
c_2=1-\frac{1}{Q_i}\,\int_{r_*}^\infty \frac{dr}{u}\,.~~~
\ee
The immediate observation is that all the coefficients are real. The imaginary terms in the EFT action come from the interplay between the gauge field and the IR geometry, which is consistent with the intuition that only the horizon plays the role of dissipation. In the following we will show that the third term is subleading and can be ignored in both the slow mode case ($\alpha>1$) and the IR mode case ($0\leq \alpha\ll 1$).  

The coefficient $c_2$ is tunable with respect to the parameter $\alpha$, and shows interesting properties. When $\alpha>1$ the second term in the expression of $c_2$ always dominates. As $\alpha$ decreases from $\alpha=1$, the two terms gradually become comparable, and for $\alpha\rightarrow 0$ the first term dominates. In particular, $c_2=0$ at $\alpha=0$. There are several relations needed to be stressed in the following. 

Firstly, notice that $c_2\ll \sigma$ in the slow mode scenario. This is because when $\alpha>1$
\bea
\frac{1}{Q_i}\int_{r_*}^\infty \frac{dr}{u}\gg 1
\eea
so that 
\bea
c_2\simeq \frac{1}{Q_i}\int_{r_*}^\infty \frac{dr}{u}\,.
\eea
Another fact is 
\bea
\int_{r_*}^\infty dr\frac{1}{u}\ll
\int_{r_*}^\infty dr\frac{e^{\alpha\phi_*}}{ue^{\alpha \phi}}=\int_{r_*}^\infty dr\frac{\sigma}{ue^{\alpha \phi}}\,,
\eea
which can be concluded from the rapid decay of $e^{\alpha \phi}$ away from the horizon when $\alpha>1$. As a result, $c_2\ll \sigma$ when $\alpha>1$.

Secondly, $c_1\gg c_2 \tau^{-1}$ in the slow mode case, since 
\bea
\frac{c_2\partial_0}{c_1}\propto
\frac{c_2\tau^{-1}}{c_1}=\frac{c_2}{c_1\tau}\propto \frac{c_2}{\sigma}\ll 1\,,
\eea
where we have used 
\bea
\label{eq:c1relation}
\frac{1}{c_1}=Q_i=\int_{r_*}^\infty dr \frac{1}{ue^{\alpha\phi}}\simeq \frac{\tau}{\sigma}\,.
\eea
The last approximate equality is based on the following observation. The integral can be separated into a finite term and a logarithmic term. In the overlap region $\omega \ll r_*-r_h\ll T$, the finite term dominates and it is nothing but $\tau/\sigma$. 

Finally, there is an important scaling property for $c_1$ in the IR mode case. When $0\leq \alpha\ll 1$ at low temperature $T$, the integral starts from $r_*$ to the UV boundary, where $\omega, T\ll r_*-r_0\ll m$. To obtain the scaling property of this integral in $Q_i$, we first take $r_0\rightarrow 0$ in $u$ and $\phi$ to have 
\bea
u(r)\simeq r^{\frac{3}{2}}\frac{\sqrt{2}m+r}{\sqrt{\frac{m}{\sqrt{2}}+r}}\,, \quad
e^{\alpha\phi}\simeq\left(\frac{\frac{m}{\sqrt{2}}+r}{r}\right)^{\frac{\alpha}{2}}
\eea
with $\omega, T\ll r_*\ll m$
and we can choose the indefinite integral to be
\bea
\begin{split}
\int dr\frac{1}{ue^{\alpha\phi}}&\simeq \frac{C_{\alpha}}{m}\left(\frac{\sqrt{2}m+2r}{2m+\sqrt{2}r}\right)^{\frac{3-\alpha}{2}}~_2F_1\left(\frac{3-\alpha}{2},\frac{3-\alpha}{2},\frac{5-\alpha}{2},\frac{2(m+\sqrt{2}r)}{2m+\sqrt{2}r}\right)\\
&\equiv \frac{C_{\alpha}}{m}\,I(\alpha,m;r)
\end{split}
\eea
as one of the possible solutions, with $C_{\alpha}$ a complex constant that only depends on $\alpha$. 
Asymptotic to the boundary $r\rightarrow \infty$, because $r\gg m$, the quantity $I(\alpha,m;r)$ reduces to a finite constant 
that is independent of $m$. Since $r_*\ll m$, we can set $r_*=\lambda m$ with $\lambda$ an arbitrary constant provided $\lambda\ll 1$. Then this quantity $I(\alpha,m;r)$ reduces to another constant 
depending on $\lambda$ and $\alpha$, which has nothing to do with the scaling behavior. 
Therefore, we conclude that 
\bea
\label{eq:c1IR}
Q_i\propto m^{-1}\,,\quad \text{or equivalently}\quad c_1\propto m
\eea
for $0\leq \alpha\ll 1$ at low temperature. We have also confirmed on this relation from numerical computations, where the numerical results show exactly the same scaling behavior in $Q_i$. We can obtain a deduction from Eq.\eqref{eq:c1IR} that $c_1\gg c_2\omega$, because \bea
c_2=1-\frac{Q_i|_{\alpha=0}}{Q_i|_{\alpha}}\sim \mathcal{O}(m^0)\,\quad\text{for a fixed}~\alpha\,,
\eea
while $c_1\propto m\gg \omega$ in the IR mode case. Therefore, the third term in the action Eq.\eqref{EFTC1C3} is negligible compared to the second term in the IR mode scenario, which is similar in the slow mode case. 

Now let us focus on the contour $\mathcal{C}_2$, i.e. the near horizon region, to calculate $S_{\mathcal{C}_2}[b_{1\mu}, b_{2\mu}]$. It turns out the action depend on the parameter $\alpha$ which will be considered separately in the following.  

\subsubsection{EFT with a slow mode}
\label{sec:slowmode}
In this subsection, we consider the EFT of the diffusive hydrodynamics in presence a long-lived slow mode. 
Compared to the other degrees of freedom with higher energy, the energy scale of these two dynamical quantities are much lower, therefore, the other degrees of freedom can be integrated out to formulate a low-energy EFT including only these two modes. 
This picture works for our system when $\alpha>1$ \cite{Liu:2021qmt}. 
In this case in addition to a hydrodynamic diffusive mode, there is a slow mode with long lifetime scaling as $\tau \sim 1/T^\alpha\gg 1/T$ at low temperature. 
This can be viewed as an example of the quasihydrodynamics \cite{Grozdanov:2018fic}, in which the conserved hydrodynamic quantities couple to approximately conserved quantities, as discussed in the appendix of \cite{Liu:2021qmt}. 
Note that the expression of $\tau$ is an integration along the radial direction and therefore it encodes the information of all energy scales \cite{Liu:2021qmt}.  

When $\alpha>1$, the low-energy dynamics of this system can be effectively described by the  telegrapher equation that can be obtained from the near-far matching method from holography \cite{Liu:2021qmt}. Therefore, we choose the end points ($r_*\pm i\epsilon$) of the contour $\mathcal{C}_2$ to be in the region $\omega, k \ll r_*-r_0\ll T$. 
The near horizon geometry and the behavior of the dilaton in the limit $r-r_0\ll T$ are 
\be
u=4\pi T (r-r_0)+\dots\,, ~~~~~
f=\sqrt{2} m \pi T+\dots\,,~~~~~
e^\phi=\frac{m}{2\sqrt{2}\pi T}+\dots\,,
\ee
where the dots represent the higher order corrections in $(r-r_0)$ and are irrelevant. 

In order to solve the EOM \eqref{defPi}-\eqref{eomPi} in the contour  $\mathcal{C}_2$, we impose Dirichlet boundary conditions for the gauge field at $r_*\pm i\epsilon$
\bea
\begin{split}
C_\mu^{(0)} (r\to r_*-i\epsilon)=b_{1 \mu}(x)\,,~~~C_\mu^{(0)} (r\to r_*+i\epsilon)=b_{2\mu }(x)\,,
\end{split}
\eea
and we will make clear the roles of $b_{1\mu}$ and $b_{2\mu}$ after obtaining the EFT action. In addition, the time component of the gauge field need to vanish at the horizon, i.e., the boundary condition \eqref{eq:c0bnd} 
since we want to have independent $b_{10}$ and $b_{20}$ (for more explanations, see \cite{Glorioso:2018mmw}). 

The solution of $C_0$ to the zeroth order is
\be
C_0^{(0)}=
 \begin{cases}
   -\displaystyle\frac{b_{10}}{q_0} \int_r^{r_*}\frac{d\tilde{r}}{f\, e^{\alpha\phi} }+ b_{10}(x)=b_{10}\,\frac{r-r_0}{r_*-r_0} +\dots  \,~~~\text{upper~} 
   \\[10pt]
   -\displaystyle\frac{b_{20}}{q_0} \int_r^{r_*}\frac{d\tilde{r}}{f\, e^{\alpha\phi} }+ b_{20}(x) =  
   b_{20}\,\frac{r-r_0}{r_*-r_0} +\dots
   \,~~~\text{lower~} 
  \end{cases}
\ee
where the integration constant
\be
\label{defq0}
q_0=\int_{r_h}^{r_*}\frac{dr}{f\, e^{\alpha\phi} }
=\frac{2^\alpha (\sqrt{2}\pi T)^{\alpha-1}}{m^{\alpha+1}}\,(r_*-r_0)+\dots\,
\ee
depends on the choice of the IR cutoff $r_*$. However, we will see that the EFT action is independent of $r_*$ in our final results. 
 
The solution of $C_i$ to the zeroth order is 
\be
C_i^{(0)}=
\frac{b_{1i}-b_{2i}}{2\pi i  }\,\text{Log}\frac{r-r_0}{r_*-r_0}+b_{2i}\,,
\ee
where $\text{Log}$ is the principle value of $\log$. 

The higher order equations are Eq.\eqref{Eqhigher}-Eq.\eqref{defsource}. Based on the zeroth order solutions $C_0^{(0)}$ and $C_i^{(0)}$, we solve these equations to the first order 
\bea
\begin{split}
C_0^{(1)}&=\int^r_{r_*+i\epsilon}dr'\frac{1}{fe^{\alpha\phi}}\left(\int^{r'}_{r_*+i\epsilon}dr''\,s_0^{(1)}+c_0^{(1)}\right)\,,
\\[7pt]
~~ C_i^{(1)}&=\int^r_{r_*+i\epsilon}dr'\frac{1}{ue^{\alpha\phi}}\left(\int^{r'}_{r_*+i\epsilon}dr''\,s_i^{(1)}+c_i^{(1)}\right)\,,
\end{split}
\eea
where 
\bea
\begin{split}
c_0^{(1)}&=\frac{1}{q_0}\int_{r_*}^{r_0}dr'\frac{1}{fe^{\alpha\phi}}\int^{r'}_{r_*}dr''\,s_0^{(1)}\,
=\frac{1}{q_0}\frac{\partial_i(b_{2i}-b_{1i})}{2\pi i}\int_{r_*}^{r_h}dr'\frac{1}{fe^{\alpha\phi}}\int^{r'}_{r_*}dr''\,\frac{1}{r''-r_0}\,,\\[9pt]
c_i^{(1)}&=\frac{1}{q_i}\int_{r_*+i\epsilon}^{r_*-i\epsilon}dr'\frac{1}{ue^{\alpha\phi}}\int_{r_*+i\epsilon}^{r'}dr''s_i^{(1)}\\[8pt]
&=\frac{1}{q_i}\frac{\partial_0(b_{2i}-b_{1i})}{2\pi i}\int_{r_*+i\epsilon}^{r_*-i\epsilon}dr'\frac{1}{u}\int_{r_*+i\epsilon}^{r'}dr''\left(-\frac{1}{r''-r_0}\right)\\[7pt]
&~~~~-\frac{1}{q_i}\int_{r_*+i\epsilon}^{r_*-i\epsilon}dr'\frac{1}{u}\,\Big[(\partial_0C_i^{(0)}-\partial_iC_0^{(0)})-(\partial_0 b_{1i}-\partial_i b_{10})\Big]\,,
\end{split}
\eea
with $q_0$ defined in Eq.\eqref{defq0} and 
$
q_i=2iT\sigma\,
$.

Let us consider $-e^{\alpha\phi}u\,C_i\partial_rC_i$ to $\partial_0$: 
\bea
-e^{\alpha\phi}u\,C_i\partial_rC_i=
\begin{cases}
 -b_{2i}c_i^{(1)}\,, \quad ~~~~~~~~~~~~~~~~~~~~~~~~~ \text{at $r_*+i\epsilon$}  \\[8pt]
 -b_{1i}\left(\displaystyle\int_{r_*+i\epsilon}^{r_*-i\epsilon}dr\,s_i^{(1)}+c_i^{(1)}\right)\,,\quad \text{at $r_*-i\epsilon$}
\end{cases}
\eea

Up to the $\partial_0^{(1)}$ order, we have
\bea\label{eq:actionC2}
\begin{split}
S_{\mathcal{C}_2}[b_{1\mu}, b_{2\mu}]
&=\int dv dx dy\,\Big[
i\sigma T\, b_{ai}b_{ai}+\sqrt{2}m\pi T\sigma \,\frac{b_{a0}b_{r0}}{r_*-r_0}
- \sigma\,  b_{ai}\partial_0b_{ri} \Big].
\end{split}
\eea

The calculation above works only for the case of the slow mode with $\tau^{-1} \ll \omega \ll T$, i.e. $\alpha>1$.  When $\alpha<1$, we have $\tau^{-1}  \sim T \ll \omega \ll m$, which we need to consider separately. 

Combing the action \eqref{EFTC1C3} on $\mathcal{C}_1\cup \mathcal{C}_3$ and the action \eqref{eq:onshellS} on $\mathcal{C}_2$,  
the total on-shell action 
can be written as 
\bea
\begin{split}
S&= \int dv dx dy\,\Big[i\sigma T\, b_{ai}b_{ai}+\sqrt{2}m\pi T\sigma \,\frac{b_{a0}b_{r0}}{r_*-r_0}- \sigma\,  b_{ai}\partial_0b_{ri}
\\ &~~~~~~~~~~~~~~~~~
+
c_0 (B_{a0}-b_{a0})(B_{r0}-b_{r0})
- c_1 (B_{ai}-b_{ai})(B_{ri}-b_{ri})\\
&~~~~~~~~~~~~~~~~~~
+ 
c_2 (b_{ai}\partial_0 B_{ri}-B_{ai}\partial_0 b_{ri})
\Big].
\end{split}
\eea

Note that when $r_*\to r_0$, we can integrate out the $b_0$ fields. The strategy is as follows. Following the mean field approximation, one first obtains the EOMs for $b_{a0}$ and $b_{r0}$ and then plugs these EOMs into the action above. Using the relations between the coefficients in the EFT action,  
\be
\label{eq:relcoeff}
c_1\simeq\frac{\sigma}{\tau}\,,
~~c_2\ll\sigma\,,
~~ c_1\gg c_2\partial_0\,,
\ee
we obtain the effective action 
\bea
\label{eq:action1}
\begin{split}
S&= \int dv dx dy\,\Big[\chi B_{a0}B_{r0}+i\sigma T\, b_{ai}b_{ai}\\ &~~~~~~~~~~~~~~~~~~- \sigma\,  b_{ai}\partial_0b_{ri}
- c_1 (B_{ai}-b_{ai})(B_{ri}-b_{ri})
\Big],
\end{split}
\eea
where we have ignored the term $c_2 (b_{ai}\partial_0 B_{ri}-B_{ai}\partial_0 b_{ri})$ 
due to the relation \eqref{eq:relcoeff},  
which could play roles when we consider physics of higher energy scale of order $c_1/c_2$. In this case we should perform the derivative expansion up to even higher order terms. 

We can perform some consistent checks for the action above. 
\begin{itemize}
    \item {\bf Recover the standard diffusion theory.}~ 
    When we integrated out the fields $b_0$ and $b_i$, we could reproduce the effective action for the U(1) charge diffusion in \cite{Crossley:2015evo}. 
From $\frac{\delta S}{\delta b_{ai}}=\frac{\delta S}{\delta b_{ri}}=0$, we have 
\bea
\label{eq:onshell}
\begin{split}
\bar{b}_{ai}=&\frac{c_1
}{c_1- \sigma \partial_0} B_{ai}\\
\bar{b}_{ri}=&\frac{c_1
}{c_1+ \sigma\partial_0} B_{ri}-\frac{2 i\sigma T c_1
}{(c_1+ \sigma\partial_0)(c_1- \sigma\partial_0)}B_{ai}
\end{split}
\eea

and taking the hydrodynamic limit
$\partial_0\sim \omega\rightarrow 0$,
we find that
\bea
\begin{split}
\bar{b}_{ai}=&\, B_{ai}\,,\\
\bar{b}_{ri}=&\, B_{ri}-\frac{2 i\sigma T }{c_1}B_{ai}\,.
\end{split}
\eea
In the second equation, $B_{ai}$ is still there but it contributes only up to total derivatives to the action. Plugging \eqref{eq:onshell} into the action \eqref{eq:action1}, we have 
\be
\mathcal{L}=\chi B_{a0}B_{r0}- \sigma B_{ai}\partial_0 B_{ri}+i\sigma T B_{ai} B_{ai}.
\ee

\item {\bf Recover the Maxwell-Cattaneo model.}~ We can compare the action \eqref{eq:action1}  with the result in  \cite{Jain:2023obu} as follows.\footnote{Here, the comparison between the EFT derived from holography for $\alpha>1$ at low temperature and the MC model is performed only up to the first order in derivatives. It would be intriguing to extend this analysis by calculating the higher-order terms in holography and comparing with the MC model incorporating generic higher derivative terms. The latter theory could be constructed using the method in  \cite{Gouteraux:2023uff}.}  
Redefine 
\be
b_{ai}=B_{ai}+v_{ai}\,,~~~~
b_{ri}=B_{ri}+v_{ri}\,,~~~~
\ee
and we have 
\be
\label{eq:action-kovtun}
\mathcal{L}=\chi B_{a0}B_{r0}+i\sigma T (B_{ai}+v_{ai})(B_{ai}+v_{ai})-c_1 v_{ai} v_{ri}-\sigma (B_{ai}+v_{ai})\partial_0 (B_{ri}+v_{ri}),
\ee
which produces the action obtained in \cite{Jain:2023obu} where one sets $
\alpha_v=\chi_v=c_1$ in Eq.\eqref{eq:action-Kovtun}, where $c_1=\sigma/\tau$ as we have discussed around Eq.\eqref{eq:c1relation}. The symmetric properties of these two fields are the same as the ones discussed in Sec. \ref{sec2}. From holography, the dynamical field $v_i$ in MC model can be understood as the difference between the gauge field at the boundary $\infty$ and $r_*$ in the far region, i.e. the contour $\mathcal{C}_1$ and $\mathcal{C}_3$ in Fig. \ref{fig:contour}. This reminds us the normal mode in a domain wall geometry for the gauge field under the Dirichlet boundary condition at the wall.\footnote{We thank Yanyan Bu for useful discussions on this point.} 

The extra vector field $v_{ri}$ plays the role of a  relaxed spatial current that is conjugate to $v_{ai}$. Apart from diffusion induced from gradient of the charge density, the current also relaxes with $\tau$ being the relaxation time. 

The constitutive equation for the 
$U(1)$ current is 
\bea
\label{eq:slowcurrent}
\begin{split}
J_{r0}&=\frac{\delta S}{\delta A_{a0}}
=\chi B_{r0}\,,\\[8pt]
J_{ri}&=\frac{\delta S}{\delta A_{ai}}
=2 i\sigma T (B_{ai}+v_{ai})-\sigma\partial_0 (B_{ri}+ v_{ri})\,,
\end{split}
\eea
where one could identify $\mu=B_{r0}$. 
The fields $v_{ai}$ and $v_{ri}$ are from
\bea
\label{eq:eomvavrslow}
\begin{split}
\frac{\delta S}{\delta v_{ai}}&=
2 i\sigma T (B_{ai}+v_{ai})-
c_1 v_{ri}-\sigma\left(\partial_0 B_{ri}+\partial_0v_{ri}\right)=0\,,\\[8pt]
\frac{\delta S}{\delta v_{ri}}&=
-
c_1 v_{ai}+\sigma\partial_0\left( B_{ai}+v_{ai}\right)=0\,.
\end{split}
\eea
Using the first equation above, we can rewrite Eq.\eqref{eq:slowcurrent} as 
\bea\label{eq:currentslow}
J_{ri}=c_1 v_{ri} \,,
\eea
which is precisely Eq.\eqref{eq:currentKovtun}, 
once we identify $c_1=\alpha_v$. The constitutive equation \eqref{eq:constiKovtun} can be identified as the first equation in \eqref{eq:eomvavrslow}. 

\item  
{\bf Green's function.}~ To study the Green's function, it is simpler to work in the momentum space where the symmetric and retarded Green's functions can be obtain as follows. We consider the fields of the form $e^{-i\omega v+ik x}$, i.e. with the momentum along the $k_x$ direction. We first integrate out the fields $v_{ai}, v_{ri}$ by plugging in their EOM and then integrate out the fields $\varphi_a, \varphi_r$ also by plugging in the related EOM, then after a straightforward calculation we obtain 
\bea
\begin{split}
G^S_{tt}(\omega, k)&=\frac{-i\, \delta^2 W}{\delta A_{at}(-\omega,-k)\,\delta A_{at}(\omega, k)}=\frac{2i\sigma T k^2}{|i\omega(1-i\omega\tau)-D k^2|^2}\,,\\[9pt]
G^S_{xx}(\omega, k)&=\frac{-i\, \delta^2 W}{\delta A_{ax}(-\omega,-k)\,\delta A_{ax}(\omega,k)}=\frac{2i\sigma T\omega^2}{|i\omega(1-i\omega\tau)-D k^2|^2}\,,\\[9pt]
G^S_{yy}(\omega, k)&=\frac{-i\, \delta^2 W}{\delta A_{ay}(-\omega,-k)\,\delta A_{ay}(\omega,k)}=\frac{2i\sigma T}{|1-i\omega\tau|^2}\,,\\
 \end{split}
\eea
and 
\bea
\label{eq:retGslow}
\begin{split}
G^R_{tt}(\omega, k)&=\frac{-i\, \delta^2 W}{\delta A_{at}(-\omega, -k)\,\delta A_{rt}(\omega, k)}=\frac{-k^2\sigma }{i\omega(1-i\omega\tau)-D k^2}\,,\\[9pt]
G^R_{xx}(\omega, k)&=\frac{-i\, \delta^2 W}{\delta A_{ax}(-\omega,-k)\,\delta A_{rx}(\omega, k)}=\frac{-\omega^2\sigma }{i\omega(1-i\omega\tau)-Dk^2}\,,\\[9pt]
G^R_{yy}(\omega, k)&=\frac{-i\, \delta^2 W}{\delta A_{ay}(-\omega,-k)\,\delta A_{ry}(\omega,k)}=\frac{i\omega\sigma }{1-i\omega\tau}\,,
 \end{split}
\eea
while all the other correlators vanish. From the above relations, it is easy to check that they satisfy the KMS condition 
\be
G_{\mu\nu}^S(\omega, k)=i\coth\big(\frac{\beta\omega}{2}\big)\, \text{Im} G_{\mu\nu}^R(\omega, k)
\ee
at the leading order in $\omega$ and we have used the relations $c_1=\sigma/\tau$ and $D=\sigma/\chi$.

Note that in the near horizon region $\mathcal{C}_2$,   from \eqref{eq:actionC2} the dual IR current of the field also satisfies the KMS condition. 

\item  {\bf Dispersion relations.}~ From the retarded Green's function \eqref{eq:retGslow}, we obtain the dispersion relation of the longitudinal modes, which satisfies the semi-circle law, i.e. 
\be
i\omega(1-i\omega\tau)-D k^2=0\,.
\ee
When $k=0$, we have poles located at $\omega=0$ and $-i\Gamma$ with $\Gamma=1/\tau$. There is no diffusive mode for the transverse fluctuations. 

\end{itemize}

\subsubsection{EFT with an IR mode}
In this subsection, we consider the EFT of the diffusive hydrodynamics in presence an IR  mode. 
When $\alpha<1$, we choose $\mathcal{C}_2$ to be the region with the cutoff satisfying $\omega, k, T\ll r_*-r_0\ll m$. In this case, in addition to a diffusive hydrodynamic mode, the lowest non-hydrodynamic mode  is an IR mode with $\tau \sim 1/T$. 
In this region we cannot peform derivative expansions in terms of $\partial_0$ any more. Nevertheless, we demonstrate that (1) the KMS relation for the IR currents; (2) $\langle j^i j^j\rangle_R\sim \mathcal{G}_\text{IR}(\omega, k)$;
(3) the boundary condition \eqref{eq:c0bnd} for $C_0$. The solution of $C_0$ in the previous subsection still applies here which gives the same action of $b_{r0}b_{a0}$ term. The IR effective action can be written in the momentum space. Here we only consider the fluctuation along $k_x$ direction, i.e. the longitudinal sector. We can write down the action 
\bea
\label{eq:c2action}
\begin{split}
S_{\mathcal{C}_2}[b_{1\mu}, b_{2\mu}]&= \int \frac{d\omega dk_x d k_y}{(2\pi)^3}\,\Bigg[ 
-2 \pi T\textcolor{black}{^\alpha} \sigma \coth\big(\frac{\beta\omega}{2}\big)\,  \mathcal{G}_\text{IR}(\omega, k, T)\, b_{ai}(-\omega,-k)b_{ai}(\omega,k)
\\ &~
-  4\pi T\textcolor{black}{^\alpha} \sigma\, \mathcal{G}_\text{IR}(\omega, k, T)  \, b_{ai}(-\omega,-k) b_{ri}(\omega,k)
-\sqrt{2}m\pi T\sigma \,\frac{b_{r0}(-\omega,-k)b_{a0}(\omega,k)}{r_*-r_0}
\Bigg]
\end{split}
\eea
with $\mathcal{G}_\text{IR}$ the retarded Green's function in the near horizon geometry which is conformal to AdS$_2$ $\times$ R$^2$. When $\alpha\to 0$, we have  \cite{Liu:2021qmt} 
 \be
\mathcal{G}_\text{IR}(\omega, k, T)=\frac{i\beta\omega (1-\alpha)}{8\pi}\frac{\cos(\frac{\pi}{2}(\alpha-\frac{i\beta\omega}{2\pi}))\,\Gamma(\frac{\alpha-1}{2})\,\Gamma(\frac{1-\alpha}{2}+\frac{i\beta\omega}{4\pi})}{\cos(\frac{\pi}{2}(\alpha+\frac{i\beta\omega}{2\pi}))\,\Gamma(\frac{1-\alpha}{2})\,\Gamma(\frac{1+\alpha}{2}+\frac{i\beta\omega}{4\pi})} T^{1-\alpha}\,,
\ee
where the $k$ dependence is trivial when $k\ll m$.\footnote{For other value of $\alpha$ when $\alpha<1$, the IR Green's function usually has a nontrivial $k$-dependence. All the dimensional qunatities are in units of $m$.} 
From \eqref{eq:c2action} we have 
\be \langle j_i j_i \rangle_R= -4\pi T\textcolor{black}{^\alpha}\sigma\, \mathcal{G}_\text{IR}(\omega, k, T) \,,~~~~\langle j_i j_i \rangle_S= i \coth\big(\frac{\beta\omega}{2}\big)\, \textcolor{black}{T^{\alpha-1}}\, \text{Im} \mathcal{G}_\text{IR}(\omega, k, T) \ee
using the fact that $\mathcal{G}^*_\text{IR}(\omega, k,  T)=\mathcal{G}_\text{IR}(-\omega, -k, T)$. 
Note that here $b_i$ should be understood as the source for the IR currents $j_i$. This can be seen from the near horizon analysis in \cite{Liu:2021qmt} as follows. Define $R_*=R(r_*)$ as the UV cutoff of the near horizon region, we have 
\cite{Liu:2021qmt} 
\be
C_i= b_i^{(1)} R^{\frac{-1+\alpha}{2}}+\dots +b_i^{(0)}+\dots\,.
\ee
Therefore, in the matching regime, the value of the field $C_i$ is precisely the source of the IR current. 

For the transverse sector, the action for the field $b_{ay}$ and $b_{ry}$ takes the same form of the first two terms in  \eqref{eq:c2action} except that now we should use $\tilde{\mathcal{G}}_\text{IR}(\omega, k, T)$ in the expression which is the retarded Green's function of the IR current $j_y$.

Combing the on-shell action on three regions, 
now the total low energy effective action becomes
\bea
\label{eq:actionnew}
\begin{split}
&S= \int dv dx dy\,\Big[\chi B_{a0}B_{r0}-
 c_1 (B_{ai}-b_{ai})(B_{ri}-b_{ri})
+ c_2 (b_{ai}\partial_0 B_{ri}-B_{ai}\partial_0 b_{ri})
\Big]+
\\ &~
+\int \frac{d\omega dk_x d k_y}{(2\pi)^3}\,\Big[-2\pi T\textcolor{black}{^\alpha} \sigma \coth\big(\frac{\beta\omega}{2}\big)\,  \mathcal{G}_\text{IR}(\omega, k, T)\, b_{ai}b_{ai}
-  4\pi T\textcolor{black}{^\alpha} \sigma\, \mathcal{G}_\text{IR}(\omega, k, T)  \, b_{ai} b_{ri}
\Big].
\end{split}
\eea
For the transverse sector, i.e. the fields $b_{ay}$ and $b_{ry}$, we should use $\tilde{\mathcal{G}}_\text{IR}(\omega, k, T)$ in the expression. Different from the previous case, the above action \eqref{eq:actionnew} works for $\omega, k, T\ll m$. 
When $0\leq \alpha\ll 1$ we have $c_2\omega\ll c_1$ and the $c_2$-term can be ignored in \eqref{eq:actionnew} and it becomes more and more important when $\omega$ increases to $m$. 
Moreover, when $\alpha\to 0$ we have $c_2\to 0$ and we have a clear IR mode and we will focus on this case.\footnote{\label{f10}{Note that we have not included the $\partial_i$
(or higher-order) terms in the first line of \eqref{eq:actionnew}, which could, in principle, appear. These terms describe interactions between the hydrodynamic mode and the IR mode, and are of order $k/m$, a ratio significantly smaller than $1$, up to the location of the pole collision. We have verified that these terms can be safely neglected as the resulting action is sufficient to produce results consistent with the numerical results in \cite{Liu:2021qmt}, even close to the pole collision point. However, in other parameters regimes (e.g. for $\alpha\neq 0$) or in different scenarios (such as in  holographic models with AdS$_2$ black hole horizon), such terms may not be ignored, and this is beyond the scope of the current study. 
} }

Similar to the case of the slow mode, from the action above we can obtain the following information.  
\begin{itemize}
\item {\bf Comparison to the result of the slow mode.} In the low-frequency limit $\omega\ll T\ll m$, we have  $\mathcal{G}_\text{IR}(\omega)=-\frac{i\omega}{4\pi T\textcolor{black}{^\alpha}}$ up to the linear order in $\omega$. Therefore, the effective action 
\eqref{eq:c2action} contributed from the contour $\mathcal{C}_2$  reproduces the results $S_{\mathcal{C}_2}$ \eqref{eq:actionC2} in Sec. \ref{sec:slowmode}.
When we consider the limit $\partial_0\sim\omega\to 0$, 
the action \eqref{eq:actionnew} can reproduce the standard diffusive EFT \cite{Crossley:2015evo} which is the same result as the previous subsection of slow mode. 

Nevertheless, there could be some essential differences 
at low frequencies compared to the action \eqref{eq:action1} although the expressions look similar. 
First, note that the matching region is chosen differently. $\mathcal{C}_2$ is chosen to be $r_*\ll m$ here while $r_*\ll T$ for the slow mode case. This means that the two EFFs \eqref{eq:action1} and \eqref{eq:c2action} work for different energy scales. 
Secondly, when we consider the total EFT action with an IR mode, the relations in Eq. \eqref{eq:relcoeff} (except the last one) among the coefficients are not satisfied anymore. The precise relations will be discussed around \eqref{eq:rel-slow}. 
Thirdly, instead of the relaxation time $\tau$ that controls the Green's functions at finite frequencies, the behaviors of the IR Green's function  $\mathcal{G}_\text{IR}(\omega)$ determines the Green's functions at finite frequencies. 

\item {\bf A class of new effective actions.} One could view \eqref{eq:actionnew} as a new action in the limit $T\sim \omega \ll m$. This differs from the action \eqref{eq:action-kovtun} which only applies in the limit $\tau^{-1} \sim \omega \sim T^{\alpha} < T \ll m$ for $\alpha>1$. Note that a crucial difference is that in the slow mode case, there is a huge energy hierarchy between other gapped mode and the slow mode. While here the systems have lots of non-hydrodynamic modes which are of the same order as the first non-hydrodynamic pole. Nevertheless we can make a truncation at a certain value depending on the physical problem that we investigated. 

One could follow the previous subsection to perform a redefinition of fields   
\be 
b_{ai}=B_{ai}+v_{ai}\,,~~~
b_{ri}=B_{ri}+v_{ri}\,,
\ee
and the resulting action in the momentum space takes the form 
\bea
\label{eq:action-irnew}
\begin{split}
\mathcal{L}&=\chi B_{a0}B_{r0}-2\pi T\textcolor{black}{^\alpha} \sigma \coth\big(\frac{\beta\omega}{2}\big)\,  \mathcal{G}_\text{IR}(\omega, k, T) (B_{ai}+v_{ai})(B_{ai}+v_{ai})-c_1 v_{ai} v_{ri}\\
&~~~~~~-  4\pi T\textcolor{black}{^\alpha} \sigma\, \mathcal{G}_\text{IR}(\omega, k, T)  \, (B_{ai}+v_{ai})\, (B_{ri}+v_{ri})\,,
\end{split}
\eea
where each term is a function of the momentum, e.g. the first term is $\chi 
B_{a0}(-\omega, -k)B_{r0}(\omega,k)$. 
Here $v_{ai}, v_{ri}$ could be viewed as an independent dynamical vector field in the theory of diffusion. This action can be easily generalized to the case with other kinds of quantum critical points dual to systems with different near horizon scaling  geometries, e.g. the AdS$_2$ geometry or the Lifshitz geometry after one should carefully check the subtle point in footnote \ref{f10}. 


One can obtain the current by variation of the action w.r.t the external field and obtain the same result as \eqref{eq:currentslow}. However, now the constitutive equation is different which is obtained from the EOM of the variation of the action w.r.t $v_{ai}$. 

\item {\bf Green's function.}~
To study the Green's function, we work in the momentum space.
We assume the fluctuations are of the form $e^{-\omega v+i k x}$ with the momentum along the $x$-direction. 
We first integrate out the field $b$ and then integrate out the field $\varphi$ by imposing the equations of motion, and the effective action is thus only a functional of the fields $A_{a\mu}$ and $A_{r\mu}$. After a straightforward calculation, we obtain 

\bea
\begin{split}
G^S_{tt}&=\frac{-i\, \delta^2 W}{\delta A_{at}(-\omega,-k)\,\delta A_{at}(\omega, k)}=\frac{-4i\pi T\textcolor{black}{^\alpha}\sigma k^2\omega^2 \coth(\frac{\beta\omega}{2})\,\text{Im} (\mathcal{G}_\text{IR})}{|(-\omega^2+4\pi T\textcolor{black}{^\alpha} D k^2\mathcal{G}_\text{IR})
-\omega^2 4\pi T\textcolor{black}{^\alpha}\tilde{\sigma} \mathcal{G}_\text{IR}|^2}\,,
\\[9pt]
G^S_{xx}&=\frac{-i\, \delta^2 W}{\delta A_{ax}(-\omega,-k)\,\delta A_{ax}(\omega,k)}=\frac{-4i\pi T\textcolor{black}{^\alpha}\sigma \omega^4 \coth(\frac{\beta\omega}{2})\,\text{Im} (\mathcal{G}_\text{IR})}{|(-\omega^2+4\pi T\textcolor{black}{^\alpha} D k^2\mathcal{G}_\text{IR})
-\omega^2 4\pi T\textcolor{black}{^\alpha}\tilde{\sigma} \mathcal{G}_\text{IR}|^2}\,,
\\[9pt]
G^S_{yy}&=\frac{-i\, \delta^2 W}{\delta A_{ay}(-\omega,-k)\,\delta A_{ay}(\omega,k)}=\frac{-4i\pi T\textcolor{black}{^\alpha}\sigma \omega^4 \coth(\frac{\beta\omega}{2})\,\text{Im} (\mathcal{G}_\text{IR})}{|1+4\pi T\textcolor{black}{^\alpha}\tilde{\sigma} \mathcal{G}_\text{IR}|^2}\,,\\
 \end{split}
\eea
and 
\bea
\label{eq:retGir}
\begin{split}
G^R_{tt}&=\frac{-i\, \delta^2 W}{\delta A_{at}(-\omega, -k)\,\delta A_{rt}(\omega, k)}=\frac{4\pi T\textcolor{black}{^\alpha}\sigma \mathcal{G}_\text{IR}k^2 }{
(-\omega^2+4\pi T\textcolor{black}{^\alpha} D k^2\mathcal{G}_\text{IR})
-\omega^2 4\pi T\textcolor{black}{^\alpha}\tilde{\sigma} \mathcal{G}_\text{IR}}\,,\\[9pt]
G^R_{xx}&=\frac{-i\, \delta^2 W}{\delta A_{ax}(-\omega,k)\,\delta A_{rx}(\omega, k)}=\frac{ \omega^2 4\pi T\textcolor{black}{^\alpha}\sigma \mathcal{G}_\text{IR} }{
(-\omega^2+4\pi T\textcolor{black}{^\alpha} D k^2\mathcal{G}_\text{IR})
-\omega^2 4\pi T\textcolor{black}{^\alpha}\tilde{\sigma} \mathcal{G}_\text{IR}}\,,\\[9pt]
G^R_{yy}&=\frac{-i\, \delta^2 W}{\delta A_{ay}(-\omega,-k)\,\delta A_{ry}(\omega,k)}=\frac{-4\pi T\textcolor{black}{^\alpha}\sigma \tilde{\mathcal{G}}_\text{IR} }{1+4\pi T\textcolor{black}{^\alpha}\tilde{\sigma} \tilde{\mathcal{G}}_\text{IR}}\,,
 \end{split}
\eea
where $\mathcal{G}_\text{IR}$ and $\tilde{\mathcal{G}}_\text{IR}$ are $\mathcal{G}_\text{IR}(\omega,k,T)$ and $\tilde{\mathcal{G}}_\text{IR}(\omega,k,T)$ respectively,  
and $\tilde{\sigma}\equiv\frac{\sigma}{c_1}$. All other Green's functions vanish. 
Obviously, the Green's functions above  
satisfy the KMS condition 
\be
G_{\mu\nu}^S(\omega, k)=i\coth\big(\frac{\beta\omega}{2}\big)\, \text{Im} G_{\mu\nu}^R(\omega, k)\,.
\ee

\item  {\bf Dispersion relations.}~ From the retarded Green's function \eqref{eq:retGir}, we can study the pole structure. The first observation is that the poles of the IR Green's function are also QNM, as observed in \cite{Liu:2021qmt}. The diffusive pole still comes from the vanishing denominator. When $\alpha\to 0$, numerically we have checked that the IR Green's function $\mathcal{G}_{\text{IR}}(i\omega)=\frac{\omega}{4\pi T\textcolor{black}{^\alpha}}$ for $\omega/2\pi T \sim 0$ to $1$, until encountering a sharp peak located at the first IR pole. 
The third term in the denominator of $G^R_{xx}$ is negligible compared to the first term $-\omega^2$, since 
\bea
\label{eq:rel-slow}
\tilde{\sigma}\omega\propto \frac{\omega}{m}\left(\frac{m}{T}\right)^\alpha\propto \frac{\omega}{m}\left(\frac{m}{\omega}\right)^\alpha=\left(\frac{\omega}{m}\right)^{1-\alpha}\ll 1\,
\eea
for $\alpha\rightarrow 0$, where we have used the fact that $c_1\propto m$ and $\omega$ is of the same order as $T$. The third term in the denominator of $G^R_{xx}$ is also negligible compared to the second term, since
$
\frac{\omega^2\tilde{\sigma}}{Dk^2}\propto \tilde{\sigma}\omega\ll 1
$.
Therefore, the standard diffusive dispersion relation survives for the longitudinal sector. 

\end{itemize}


\section{Conclusions and discussions}
\label{sec:cd}

We have constructed the SK effective field theories from holographic Gubser-Rocha model at low temperature. By introducing an IR cutoff to divide the bulk contour into two parts, as shown in 
Fig. \ref{fig:contour}, our final action is the summation of the UV and IR parts of the on-shell action. We impose the Dirichlet boundary condition for the gauge field at the IR cutoff and interpret the gauge field as dynamical degrees of freedom associated with the gapped mode. In the case of the presence of a slow mode in addition to the diffusive hydrodynamic mode, the action is given by \eqref{eq:action1}. This result is equivalent to the action proposed in  \cite{Jain:2023obu} of the Maxwell-Cattaneo model of diffusion. In the case of the presence of an IR mode, the action is given by \eqref{eq:actionnew}
.  Although here we focused on a particular IR geometry which is conformal to AdS$_2$$\times$R$^2$, the formulae can be easily generalized to arbitrary other types of IR geometries. In both cases when we focus on physics 
at extremely low frequency or long time the standard diffusion hydrodynamics is recovered \cite{Crossley:2015evo}. Moreover, 
we have checked that the KMS conditions are satisfied and the dispersion relation reproduces the behavior observed numerically in \cite{Liu:2021qmt}.

There are a number of open questions to be investigated from the effective actions. Firstly, one could study the impact of nonhydrodynamic modes on the long time tail. It is important to understand the effect of the non-hydrodynamic vector fields on the diffusive pole and possible observable effects. Given the potential connection between hydrodynamics and strange metals in high Tc superconductors \cite{Davison:2013txa, Balm:2022bju}, we hope that the EFT approach could provide valuable insights into the underlying mechanism.  
Secondly, it is interesting to derive the SK effective action including even higher non-linear order terms. In our approach we have considered the Maxwell theory in the bulk which could only give the quadratic terms in hydrodynamic variables. One might consider non-linear Maxwell theory, like DBI to study the non-linear terms in the effective action.  
Finally, it would be interesting to generalize the derivation for diffusion to the effective action for the full dissipative hydrodynamics to derive the SK effective theory of Müller-Israel-Stewart theory and to compare with the results in \cite{Jain:2023obu, Mullins:2023ott}. 
For example, one could consider the hydrodynamic system with momentum weakly broken to get the SK  effective action. In this case, different from the linear Maxwell theory, we need to deal with the full nonlinear gravitational theory.  

\vspace{.3cm}
\subsection*{Acknowledgments}
 We thank Matteo Baggioli, Yanyan Bu for extensive  discussions on related questions. This work is supported by the National Natural Science Foundation of China grant No. 12375041 and No. 12035016. X.-M.W. is supported by the China Postdoctoral Science Foundation (Grant No. 2023M742296).

\end{document}

\textcolor{red}{Plugging in the zeroth order solution, we obtain\comment{CHECK: fix coefficients}
\bea
\begin{split}
S_{\mathcal{C}_1\cup\mathcal{C}_3}&=-\frac{1}{2}\int dvdxdy \bigg[
\frac{1}{Q_{i}} (B_{1i}-b_{1i})^2-\frac{1}{Q_{0}} (B_{10}-b_{10})^2-\bigg(
\frac{1}{Q_{i}} (B_{2i}-b_{2i})^2
\\&~~~~~~~
-\frac{1}{Q_{0}} (B_{20}-b_{20})^2
\bigg)
\textcolor{red}{+
(B_{1i}
-b_{1i}
)\partial_0 (B_{1i}
-b_{1i}
)-(B_{2i}
-b_{2i}
)\partial_0 (B_{2i}
-b_{2i})}+
\\
&~~~~~~~~
+
(B_{1i}
-b_{1i}
) (\partial_0 b_{1i}
- \partial_i b_{10}
)-(B_{2i}
-b_{2i}
)(\partial_0 b_{2i}
- \partial_i b_{20}
)
\\
&~~~~+
(B_{1i}
-b_{1i}
)\partial_i (B_{10}
-b_{10}
)-(B_{2i}
-b_{2i}
)\partial_i (B_{20}
-b_{20})+
+\dots +F_{ij}F^{ij}\bigg]
\\&
=
c_0 (B_{a0}-b_{a0})(B_{r0}-b_{r0})
- c_1 \mathcal{B}_{ai}\mathcal{B}_{ri}
+ 
c_2 (B_{ai}\partial_0 b_{ri}+B_{ri}\partial_0 b_{ai})
\textcolor{red}{+ c_3 (\mathcal{B}_{ai}\partial_i \mathcal{B}_{r0}+\mathcal{B}_{ri}\partial_i \mathcal{B}_{a0})}
\\&~~~~~~~~+c_4\,F_{ij}F^{ij}
\end{split}
\eea}
where 
\be
c_0=\frac{1}{Q_0}=\chi\,, ~~ c_1=\frac{1}{Q_i}\,,~~~
c_2=\,,~~~
c_3=\,,~~~
c_4=\,,
\ee